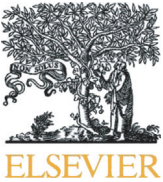
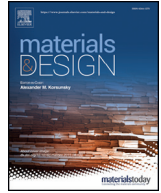

# Fracture mechanics of micro samples: Fundamental considerations

R. Pippan [a], S. Wurster [a],*, D. Kiener [b]

[a] Erich Schmid Institute of Materials Science, Austrian Academy of Sciences, 8700 Leoben, Austria
[b] Department Materials Physics, Montanuniversität Leoben, 8700 Leoben, Austria

## HIGHLIGHTS

- Fundamental concepts of fracture mechanics are revisited with respect to their applicability in miniaturized experiments
- Stress state and notch sharpness in miniaturized samples need to be considered
- For size independent fracture toughness values, sample sizes must obey relations to the microstructure and fracture process zone
- Different contributions to material toughness can be separated by appropriate experimental design

## GRAPHICAL ABSTRACT

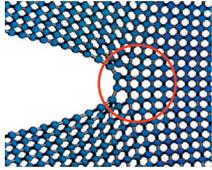

## ARTICLE INFO

## ABSTRACT

In this review article we consider the crack growth resistance of micrometer and sub-micrometer sized samples from the fracture mechanics point of view. Standard fracture mechanics test procedures were developed for macro-scale samples, and reduction of the specimen dimensions by three to five orders of magnitude has severe consequences. This concerns the interpretation of results obtained by micro- and nano-mechanics, as well as the life time and failure prediction of micro- and nano-devices. We discuss the relevant fracture mechanics length scales and their relation to the material-specific structural lengths in order to conduct rigorous fracture mechanics experiments. To ensure general validity and applicability of evaluation concepts, these scaling considerations are detailed for ideally brittle, semi-brittle and micro ductile crack propagation, subject to both monotonic and cyclic loading. Special attention is devoted to the requirements for determining specimen size for various loading types to measure material characteristic crack propagation resistance at small scales. Finally, we discuss novel possibilities of micron and sub-micron fracture mechanics tests to improve the basic understanding of specific crack propagation processes.

## 1. Introduction

Fracture mechanics was first proposed about 100 years ago. Griffith [1] used an energy analysis to determine the conditions for the propagation of pre-existing cracks in ideally brittle materials. Since this idea was not applicable to ductile materials such as steels or aluminum alloys, it was not utilized in engineering applications until 1948 when Irwin [2] and Orowan [3] extended the Griffith approach to metals by including the energy dissipation arising from local plastic flow. Irwin [2] introduced the terms energy release rate and stress intensity factor. These quantities are now used as crack driving forces under small scale yielding conditions, and the regime where the use of these parameters is valid is denoted as linear elastic fracture mechanics (LEFM).

In the 1960s, Paris [4] provided convincing experimental results that suggested fracture mechanics can be extended to fatigue by using the

* Corresponding author.
 E-mail address: stefan.wurster@oeaw.ac.at (S. Wurster).

**Nomenclature**

| | |
|---|---|
| a | Crack length |
| B | Specimen thickness |
| b | Length of burgers vector |
| CTOD, $CTOD_c$ | Crack tip opening displacement, critical |
| da/dN | Cyclic crack growth rate |
| DFZ | Dislocation free zone |
| E | Young's modulus |
| EPFM | Elastic plastic fracture mechanics |
| $f_{ij}(\theta)$ | Angular dependent geometry factor |
| FIB | Focused ion beam |
| G | Energy release rate |
| J, $J_i$, $J_c$ | J-integral, initiation of propagation, critical |
| K, $K_I$, $K_{ic}$ | Stress intensity factor, loading mode I, II, III, critical |
| LEFM | Linear elastic fracture mechanics |
| MEMS | Micro electro mechanical system |
| n | Strain hardening exponent |
| NEMS | Nano electro mechanical system |
| R | Crack resistance |
| $r_K$ | Radius of the K-dominated zone |
| $r_{pl}$ | Radius of the plastic zone |
| $r_{HRR}$ | Radius of the HRR-field dominated zone |
| $r_{fr}$ | Radius of the fracture process zone |
| SEM | Scanning electron microscope |
| TEM | Transmission electron microscope |
| W | Specimen width |
| W-a | Ligament |
| $x_i$ | Space coordinates in direction i |
| $\gamma_0$ | Surface energy |
| $\gamma_{pl}$ | Plastic energy |
| $\gamma_{pl\ fr}$ | Plastic energy of for generation of fracture surface |
| $\gamma_{pl\ plz}$ | Plastic energy to advance plastic zone |
| $\Delta CTOD$ | Cyclic crack tip opening displacement |
| $\Delta J_{eff}$ | Effective cyclic J integral |
| $\Delta J_{th}$ | Threshold of the cyclic J integral |
| $\Delta K_{eff}$ | Effective stress intensity factor range |
| $\Delta K_{th}$ | Cyclic threshold stress intensity factor range |
| $\Delta r_{fr}$ | Radius of the cyclic fracture process zone |
| $\Delta r_{pl}$ | Radius of the cyclic plastic zone |
| $\mu$ | Shear modulus |
| $\sigma_0$ | Reference stress |
| $\sigma_{ap}$ | Applied stress |
| $\sigma_{ij}(r,\theta)$ | Stress component ij dependent on radius r and angle $\theta$ |
| $\sigma_t$ | Theoretical strength |
| $\sigma_y$ | Yield stress |
| $\tau_{r\theta}$ | Radius and angular dependent shear stress |
| $\nu$ | Poisson's ratio |

stress intensity factor range as a crack driving force. Wells [5] and Rice [6] developed the parameters crack tip opening displacement (CTOD) and J-integral, respectively, which permit the description of crack propagation for materials or structures that are too ductile for LEFM consideration. These now serve as loading parameters for elasto-plastic fracture mechanics (EPFM). Hutchinson [7], Rice and Rosengren [8] related the J-integral to the crack tip stress field in non-linear materials, and Shih [9] elaborated a relation between the J-integral and CTOD, illustrating that both parameters are equally valid for characterizing fracture.

These works are the basis for test standards that have been developed for fracture toughness, fatigue crack propagation and environment assisted crack propagation, as well as for fracture mechanics based design of components – the concept of, "damage tolerant" design [10].

Common standard fracture mechanics tests are established for sample sizes ranging from about 10 mm to few 100 mm.

However, fracture mechanics is not limited to the macroscopic world. The size of devices used in micro-electro-mechanical systems (MEMS), such as sensors and actuators, microelectronics, and various medical devices has been reduced to the micron and sub-micron regime. Like macro-scale components, these devices also contain flaws that govern their fracture load, fracture strain, and lifetime [11–14]. Furthermore, macro-scale fracture mechanics analysis and test procedures are inappropriate for micro- and nano-scale components. The development of a methodology to extend fracture mechanics to micro- and nanoscale dimension is an essential task for the future.

In the last decade, significant developments have been made in the experimental determination of fracture resistance on the microscale. Most importantly for this progress was the establishment of focused ion beam (FIB) devices in material science laboratories [15–17]. FIB tools permit the machining of well-defined notched samples with dimensions of several nanometers up to several micrometers. Other machining techniques exist, such as lithography, a combination of ion slicer and FIB [18], or femto-second laser machining [19]. These techniques enable the preparation of samples with sizes that bridge the gap between typical FIB specimens and the standard macro-scale samples for conventional tests.

The potential of these new preparation techniques and related development of micro mechanical experiments for testing of miniaturized samples is well documented [17,20]. A comprehensive review about the recent development of micro mechanical experiments including fracture mechanics experiments has been published by Dehm et al. [21], and a review of nanometer-scale fracture mechanics experiments has been published by Kitamura et al. [22,23].

Since the transfer of standard fracture mechanics evaluation procedures from the macro- to micro- and nano-regime is often not straightforward, the present paper is intended to provide guidelines to interpret the results of such experiments conducted on the microscale.

The main questions addressed in this paper are:

1. What are the essential fracture mechanics length scales?
2. When can standard LEFM or EPFM be applied to micro-samples?
3. When are the fracture mechanics quantities obtained on micro-samples characteristic values independent of sample size?
4. What are the fundamental reasons for specimen size, loading type or structure dependent fracture resistance?
5. How can micro-samples contribute to improve the understanding of crack growth processes in general?

## 2. Important length scales of fracture mechanics

The stress and strain fields in front of a crack have been studied in the last century for elastic [24], elastic-plastic [7,8], and viscoelastic cases [25]. The evolution of the stress field during loading of a cracked linear elastic-plastic material is schematically illustrated in Fig. 1.

During elastic loading, the stress field in the vicinity of the crack tip is described by:

$$\sigma_{ij}(r,\theta) = \frac{K}{\sqrt{2\pi r}} f_{ij}(\theta), \tag{1}$$

where the stresses $\sigma_{ij}(r,\theta)$ are described by the stress intensity factor K, the distance from the crack tip r and an angular dependent term $f_{ij}(\theta)$ [10]. Fig. 2. illustrates the comparison of the near crack tip approximation of the normalized stresses and the normalized exact solution of the stresses [26] for a through thickness crack with length 2a in an infinite sheet in the loading ($x_2$) and the crack propagation direction ($x_1$). The region where this near tip stress field (Eq. (1)) is a good approximation for the real stress field is about 1/5 of the crack length, a, as long as this is smaller than the ligament width (W-a), where W is the specimen width. The region of this near tip field is called the K-dominated zone. If the ligament width is smaller than the crack length, the size of the K-dominated zone is a certain fraction of the ligament width, similar to the case of smaller crack length where the K dominated zone is a certain fraction of the crack length.

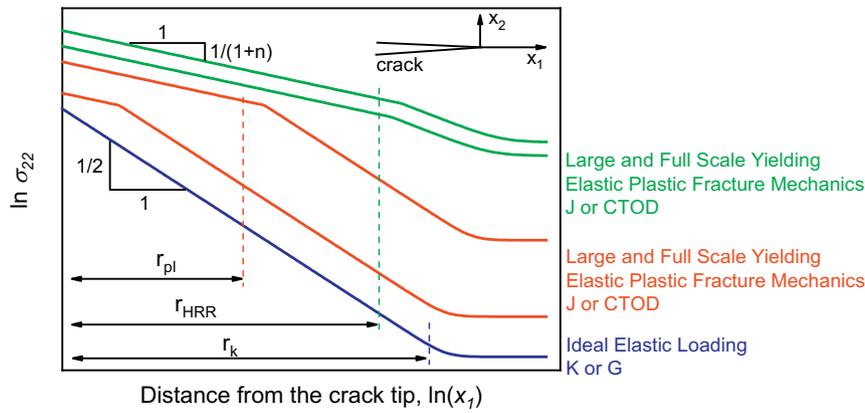

**Fig. 1.** Schematic illustration of the stress in front of a tension loaded crack along the $x_1$ axis for different loading cases from pure ideal elastic loading to full scale yielding. The different loading regimes of fracture mechanics are indicated. For large distances from the crack tip, the stress field is given by the far-field applied stress. Stresses near the crack tip are defined by Eq. (1). For strain hardening material, the near tip solution for $\sigma_{22}$ is defined by Eq. (3) with the strain hardening exponent $n$ giving the slope.

The far-field stresses to induce purely elastic loading at macroscopic cracks are quite small. Due to the stress singularity in (Eq. (1)), plastic deformation near the crack tip takes place with increasing load. The size of this plastic zone is an important length scale for fracture mechanics. Its size ($r_{pl}$) is given by the following:

$$r_{pl} = c_1 \frac{K^2}{\sigma_y^2} \quad (2)$$

and is dependent on the stress intensity factor ($K$) and yield stress ($\sigma_y$) [10]. The factor $c_1$ depends on the loading geometry, the angle in front of the crack, the stress state (plane stress, plane strain) and slightly on the hardening coefficient of the material. Its maximum value is about 0.3 [10]. As long as this plastic zone ($r_{pl}$) is embedded in the $K$ dominated zone ($r_{pl} < r_K$), the stress intensity factor and the elastic-plastic properties of the material control the fracture behavior in terms of extent of the plastic zone, deformation within the plastic zone, and plastic opening of the crack, CTOD. In fracture mechanics, this type of loading is referred to as small scale yielding, and LEFM can be applied. If the $r_{pl}$ is larger than $r_K$, EPFM is used to describe the stress and strain behavior in front of a crack. In this case, the J-integral is used as loading parameter to describe the stress and strain field of an elastic-plastic material.

The stress field in the plastically deformed region near the crack tip is described by the HRR-field after Hutchinson, Rice and Rosengren [7,8]:

$$\sigma_{ij}(r,\theta) = \sigma_0 \left(\frac{EJ}{\alpha \sigma_0^2 I_n r}\right)^{\frac{1}{1+n}} s_{ij}(n,\theta) = c_2 \left(\frac{J}{r}\right)^{\frac{1}{1+n}} s_{ij}(n,\theta) \quad (3)$$

Here, $\sigma_0$ is the reference stress of the power law approximation of the material hardening that is usually set equal to the yield stress, $E$, the Young's modulus, $n$, the strain-hardening exponent, $\alpha$, a dimensionless constant to scale the hardening, $I_n$, a constant that depends on $n$, and $s_{ij}$, a dimensionless function of $n$ and $\theta$. A material with a power law hardening exponent $n = 1$ corresponds to a linear elastic material and $n = \infty$ to an elastic-perfectly plastic material. Therefore, in Fig. 1 the slope of the stresses in the HRR-dominated regime has a $-1/(1+n)$ dependence. The size, $r_{HRR}$, where the HRR-field is a good approximation for the stress field is either (in small scale yielding) a fraction of the plastic zone size or (for large scale yielding) a fraction of the crack length. The limitation of the HRR-field in EPFM is that it is more sensitive to the type of loading (tension, bending, mixture of bending and tension) than the K field in small scale yielding, which is explained in detail elsewhere [10,27–29]. This sometimes induces even a loading type dependent fracture resistance in macrosamples and is denoted as constrain effect. Hence, a clear definition of $r_{HRR}$ is not as straightforward as for $r_K$.

Another important fracture mechanics dimension of elastic-plastic materials is the crack tip opening displacement, CTOD. In case of small scale yielding, the plastic opening is proportional to $K^2/(\sigma_y E)$ [10]. In large scale yielding, CTOD is proportional to $J/\sigma_y$ [9]. Weertman [26] refers to these length parameters as the "magic lengths of fracture mechanics" (Fig. 3), and in the case of small scale yielding they can be simply related to (K/stress)$^2$ as shown in Fig. 2.

Another important material-related length for crack growth is the size of the fracture process zone ($r_{fr}$). This length can vary over more than ten orders of magnitude, depending on the material and the fracture process. The size of this fracture process zone is important when considering the size dependence of a fracture process or the size dependence of the measured fracture resistance. The origin and nature of the fracture process zone will be elaborated in Section 4.

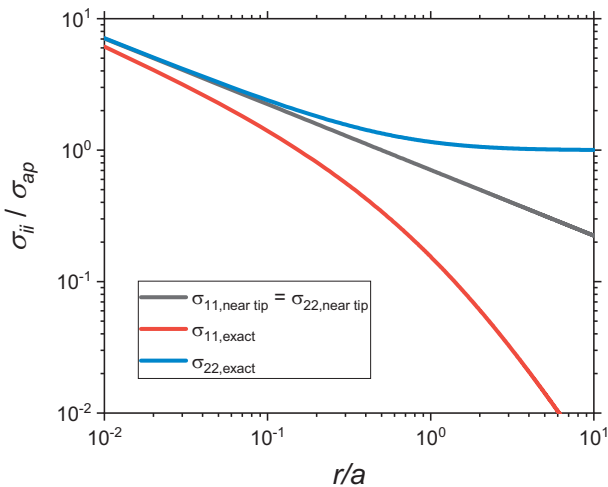

**Fig. 2.** Comparison of the near tip solution and the exact solution [26] for $\sigma_{11}$ and $\sigma_{22}$ along the $x_1$ axis for through thickness crack in an infinite sheet plotted on a log-log scale. Eq. (1) results in the solid black line with a slope of $-1/2$. The stresses are normalized by the applied stress $\sigma_{ap}$ and the distance is normalized by the crack length $a$.

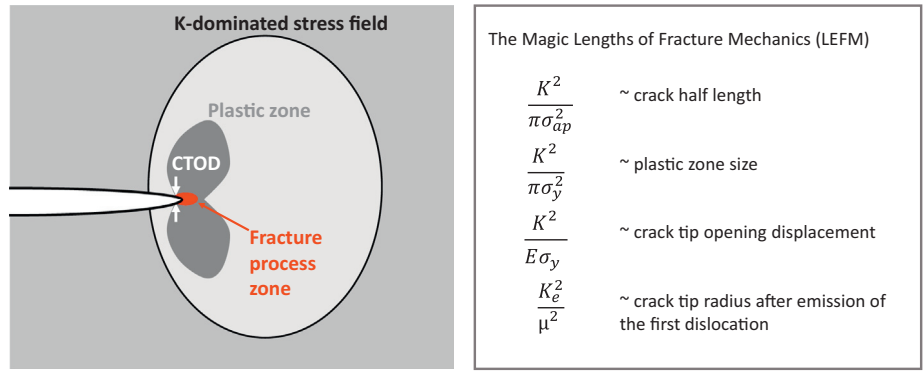

Fig. 3. Illustration of the important fracture mechanics length scales for the small scale yielding case and the summary of the "magic length scales of fracture mechanics under LEFM" after J. Weertman [26].

In order to obtain a sample size-independent measurement of crack propagation in fracture mechanics specimens, the extent of the fracture process zone should be smaller than the characteristic dimensions of fracture mechanics samples. These dimensions encompass crack length, ligament length, and sample thickness, as well as the length scale controlling the typical crack tip stress field, $r_K$ or $r_{HRR}$.

## 3. Limits to the applicability of J2-plasticity

The standard plasticity theory (J2-plasticity) is rarely applicable for describing the stress and strain distribution in front of a crack with micron dimension or for micron-sized samples, as it does not account for any microstructural influence [30]. Thus, only in nanocrystalline materials or nanocomposites where the microstructural dimensions and the dislocation spacing are significantly smaller than the typical dimensions of miniaturized fracture test samples can J2-plasticity be employed. However, preparation of microsamples from typical microcrystalline materials or microstructured composites commonly results in samples containing only a few microstructural domains, and cannot be treated using the standard J2-plasticity analysis.

Consequently, for the description of stresses and deformation in front of the crack of a typical micro-sample, the discrete nature of plasticity must be taken into account. Crystal plasticity models might be applicable to estimate the deformation behavior of a micro-sample at larger plastic deformation [31], however, one has to consider additional effects such as size and stress/strain gradient dependencies.

For very small amounts of plastic deformation, such as in the decohesion process during semi-brittle crack propagation, only few dislocations are involved. Discrete dislocation or molecular dynamics models are required to analyze the stress and deformation behavior in front of a crack where the relevant fracture processes take place [26,31–34].

Important consequences of the discrete nature of plasticity in micro fracture mechanics samples will be discussed in one of the next chapters, before considering the size of the fracture process zone. In typical standard fracture mechanics samples, the first dislocations emitted from the crack tip or generated from internal sources near the tip are embedded in the K-dominated zone. The description in the 2D case, as schematically depicted in Fig. 4, is well established [26,31,33–45]. The dislocations generated usually reduce the elastic stress field at the crack tip – a phenomenon called dislocation shielding (see Figs. 5 and 6 [31]). However, depending on their Burgers vector and position, dislocations can also increase the local elastic stress field. In such a configuration they are referred to as anti-shielding and may induce decohesion at the crack tip. The description of shielding and anti-shielding is straightforward [26,31,34] in the case of long cracks when the dislocations are located in the K-dominated zone (see Fig. 6). In case of single crystalline or bi-crystalline micro-samples, even the first dislocations generated might slip past the K-dominated zone. Hence, the crack tip shielding in micro-samples can be different than that observed in macro-scale samples. This should result in a sample size-dependent fracture resistance even in relatively brittle materials, controlled by the interplay between slip distance (at size dependent stresses), specimen dimensions, and the extent of the K-dominated zone. An important additional length scale which has to be taken into account when dislocations are generated from the crack tip is the formation of a dislocation free zone (DFZ) in front of the crack tip [26,31,34]. The size of this zone in macro-scale samples is mainly governed by the critical stress intensity to generate a dislocation at the crack tip, and the friction stress of the dislocation [46]. In micro-samples the arrangement of the dislocations and the extent of the dislocation free zone can be affected by the sample size and the loading type.

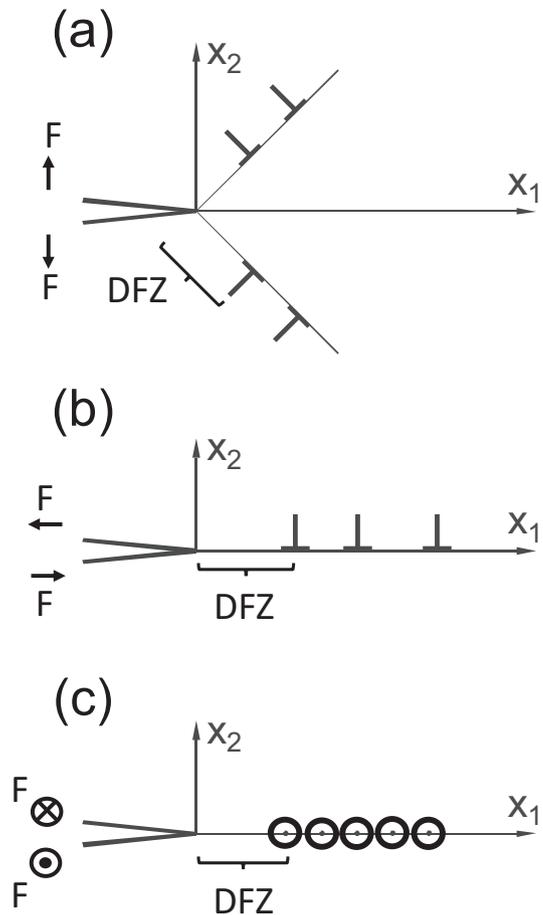

Fig. 4. Schematic illustration of the dislocation arrangement in front of a mode I (a), mode II (b), and mode III (c) loaded crack for dislocations generated at the crack tip. The applied forces, shear stresses respectively, are indicated with arrows. For cases (a) and (b), edge dislocations are present in front of the crack tip, whereas for case (c), screw dislocations are shown. The dislocations generated at the crack tip or in the immediate vicinity of the crack tip move away from the crack tip and a dislocation free zone (DFZ) develops [46].

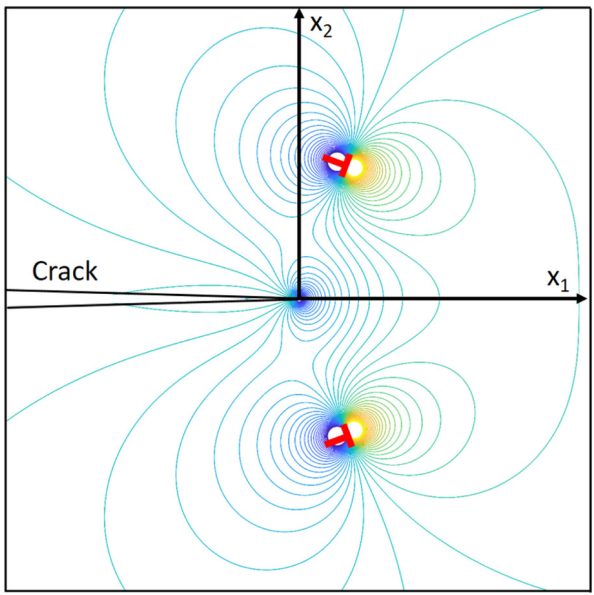

**Fig. 5.** Contour plot of $\sigma_{12} + \sigma_{22}$ induced by two symmetrically arranged dislocations (red) in front of a crack. The compression and tension regions in the surrounding of the dislocations are clearly evident. The induced compressive stresses (blue) in the immediate vicinity of the crack tip shows the typical character of the near tip stress solution known from externally loaded cracks. Tensile stresses next to the edge dislocations are shown in yellow. This illustrates the crack tip shielding from emitted dislocations. The plot is derived from complex potentials of edge dislocations and their image stress fields in presence of a crack detailed in [31].

## 4. Consideration of the fracture process zone

To discuss the dimensions of the fracture process zone it is helpful to divide materials into three classes: ideally brittle, micro ductile and semi brittle materials (see Fig. 7).

### 4.1. Ideally brittle crack propagation

If only the work of separation of atomic bonds is required for the propagation of a crack, the failure process is usually denoted as ideally brittle and the fracture resistance as the Griffith toughness. The fracture process zone of ideally brittle materials is quite small [26], on the order of 1 nm, and in a range where the non-linearity during separation of the atomic bonds occurs. In micrometer-sized ideally brittle materials the fracture process zone is usually smaller than $r_K$, and the fracture toughness values determined should not be size dependent, except when the crack length or the sample size are in the nm regime.

When the initial crack length is very short, of the order of 10 nm or below, the apparent fracture toughness will become size dependent, even for ideal brittle materials. In such a situation, the fracture process zone becomes equal to or larger than the K-dominated zone, and the fracture stress approaches the theoretical strength of the material. For this case the global fracture stress can be estimated by taking into account the nonlinearity of the stresses in the fracture process zone using a Barenblatt model [35] (see Fig. 8).

One of the practical difficulties concerning the determination of the fracture toughness of ideally brittle material is the generation of an atomically sharp pre-crack and the alignment of the cracks with respect to a crystallographic plane, a grain or phase boundary. For comparison, the typical FIB machined notch has a radius of about 10–20 nm, and can be placed with nm accuracy.

The generation of a pre-crack in ideally brittle materials is also difficult in macro-scale samples, as fatigue-induced crack propagation in ideal brittle materials does not occur. However, in micro samples, fatigue crack propagation has been reported [47]. The reason for this special phenomenon is under discussion, but it seems to be an environmental effect. Fatigue crack propagation in vacuum has not been observed. Even if such cracks can be generated by fatigue loading of the miniaturized specimen, using cyclic compression [48,49] or bending [50,51], the question remains: Is the crack plane aligned to the crack plane or interface of interest? This is also a problem in semi-brittle materials where fatigue crack initiation is often possible, but the fatigue crack plane does not coincide with the cleavage crack plane, and the crack front is not aligned with the ideal crack plane [52].

Furthermore, it is questionable whether a notch root radius of somewhat larger than 10 nm affects the measured fracture toughness of ideally brittle materials. Significantly sharper notches can be generated using the condensed electron beam of a transmission electron microscope, TEM, [53] where notch root radii of about 1 nm can be obtained. Since the microscope offers atomic resolution, precise alignment of the crack plane to a specific crystal plane is in this case also possible. However, this technique is limited to typical TEM sample thicknesses of several tens to a few hundreds of nm. Similar small notch root radii might be obtained by He-FIB [54].

For fracture toughness tests on micro-samples prepared by FIB milling, it is often assumed that irregularities along the notch front are sufficient to generate an atomically sharp pre-crack at stress concentrations along the notch at loads smaller than the global fracture load [55]. The rationale behind this is that the final sharpening of the notch is usually performed by a FIB line scan in the root of the notch in the crack propagation direction, where the local sputtering should generate such irregularities.

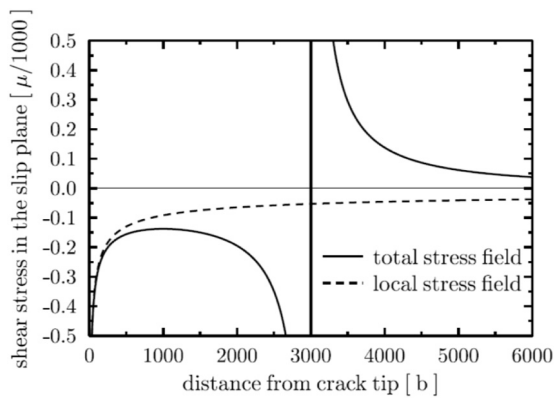

**Fig. 6.** Illustration of the shear stress in front of a mode III crack induced by a screw dislocation located 3000 Burgers vectors from the crack tip. The exact solution (solid line) is compared with the near tip solution (dashed line) taking into account only the shielding stress intensity factor of the dislocation in front of the crack tip but not the stress field of the dislocation itself. For an edge dislocation in front of a mode II crack, the evolving stress fields are very similar, except for an additional factor of $1/(1 + \nu)$ for shear stresses of edge dislocations. As for Fig. 5, complex potentials were used to calculate the stress fields [38].

| | Size of the fracture process zone |
|---|---|
| **Ideally brittle material** | |
| 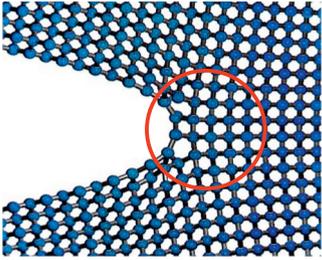 | ~ nm |
| **Semi-brittle material** | |
| 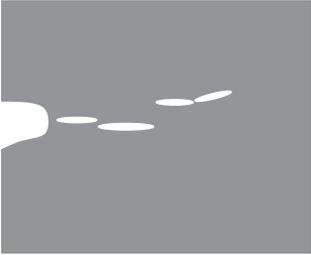 | nm – m |
| **Micro ductile material** | |
| 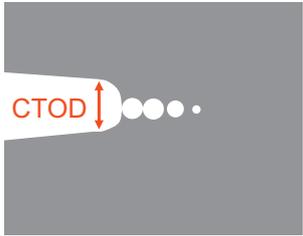 | ~ 3 CTOD$_c$<br>~ 10 nm to ~ 100 μm |

**Fig. 7.** Illustration of the size of the fracture process zone for the three different types of crack propagation processes: the ideally brittle, the semi brittle and the ductile case. The red circle gives an approximation for the size of the fracture process zone for ideally brittle materials.

In macro-scale samples, different techniques to generate an atomically sharp pre-crack have been developed, see for example [56,57]. The most universal technique is the use of special chevron ligaments, chevron-notched short rod or short bar specimens, where the stress intensity at constant load decreases to a minimum value upon crack growth, beyond which it increases as usual [58]. The advantage of this sample geometry for fracture toughness measurements is that the only parameter of import is the maximum load. Such samples are difficult to prepare on the microscale with sufficient accuracy, but notable progress was achieved recently [59–61]. However, it should be noted that the simple procedure for evaluating the fracture toughness of chevron-notched samples is only possible in materials without an R-curve behavior of the fracture resistance, so where the fracture toughness does not increase with crack extension.

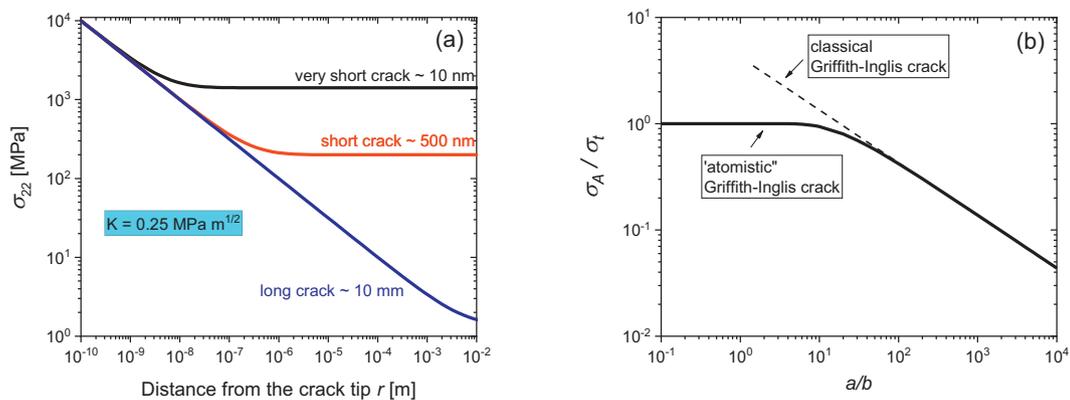

**Fig. 8.** Comparison of the stress field in front of a crack for cracks with different lengths loaded by the same stress intensity factor for ideal elastic loading and its effect on crack size dependence of the fracture strength of an ideal brittle material. (a) The fracture process zone reaches the order of the K-dominated zone when the crack length becomes smaller than 10 nm. The consequences for the fracture stress $\sigma_A$ versus crack length are shown in (b), where the Barenblatt-model is used to describe the transition to the theoretical strength $\sigma_t$. The dotted line describes the classical Griffith-Inglis crack assuming an ideal linear elastic material behavior [26].

As a simpler alternative to the generation of a chevron-like notch, remaining thin walls (see Fig. 9) can be used to permit the generation of an atomistically sharp crack at smaller loads (or smaller displacements) before unstable fracture takes place [55,62,63]. Such samples are easier to measure with displacement controlled experiments [64]. In load controlled experiments, unstable fracture may occur before the crack approaches the notch root. Additionally, the actual crack length at failure can be affected by the ligaments.

In summary, for the determination of the crack propagation resistance and fracture toughness of micro-scale ideally brittle materials, the difficulties that must be accounted for in order to determine a valid fracture toughness are similar to that of macro-scale samples.

### 4.2. Micro ductile crack propagation

Micro ductile crack propagation is the opposite extreme to brittle fracture. This fracture process can be subdivided into the following subsequent steps:

- Crack tip blunting by plastic deformation,
- Formation of micro or nano pores at inclusions, precipitations, grain or phase boundaries, or intersections of shear localizations,
- Growth of these pores,
- Coalescence of pores with the blunted crack.

Hence, the generation of the fracture surface is a purely plastic process. The dimple fracture surface often observed is formed by the plastic necking of the ligament between the pores [65].

Due to the strong strain concentration at the crack tip, the extent of the fracture process zone, where pore formation and their coalescence with the crack tip occurs, is few times CTOD [10,66] (2–5 CTOD). The CTOD value where this coalescence between pores and blunted crack tip takes place is denoted as critical crack tip opening displacement ($CTOD_c$). It can be related to the critical stress intensity or the critical J-integral under small or large scale yielding, respectively. The requirement for the $CTOD_c$ to be sample size independent is that all sample dimensions, $a$, $B$, and $W$-$a$ are at least 10–20 times CTOD [67]. In small scale yielding, CTOD is proportional to the plastic zone size, thus the fracture process zone size is proportional to $(\sigma_y\, r_{pl})/E$.

For micro-ductile materials $CTOD_c$ can vary from tens of nm to a few mm. For commonly used structural materials in safety-relevant applications, $CTOD_c$ is between 10 and 100 μm [68–70]. $CTOD_c$ values in the mm regime are typical for pure metals with a very low density of inclusions. $CTOD_c$ values smaller than 1 μm are rarely observed for materials used for structural components. These low values can mostly be found in ultrafine grained, nanocrystalline or nanocomposite structures. Micro-ductile fracture with $CTOD_c$ values below 1 μm are observed in submicron- or nanometer-thick ductile interlayers between ceramic-like materials, for example in ultrafine hard metals [71].

Hence, the measurement of sample size-independent fracture toughness values for micro-ductile materials using micro-sized specimens is limited to a small class of materials, see e.g. [72–74]. For most micro-ductile materials, fracture mechanics tests on micro-sized samples will result in a size dependence of the fracture resistance [75], as some or all requirements on $a$, $B$, and $W$-$a$ are not fulfilled.

This sample size dependence is important for the prediction of the damage tolerance of micro- and nano-electro-mechanical devices (MEMS and NEMS). These tests can also help to improve the understanding of the different phenomena taking place during micro-ductile crack propagation in macro-scale samples by elucidating details of micro-ductile fracture processes. Examples are the pore formation or the mechanical response during pore coalescence. Both could be addressed by analyzing the final necking of the ligament using samples of different size or ligament width.

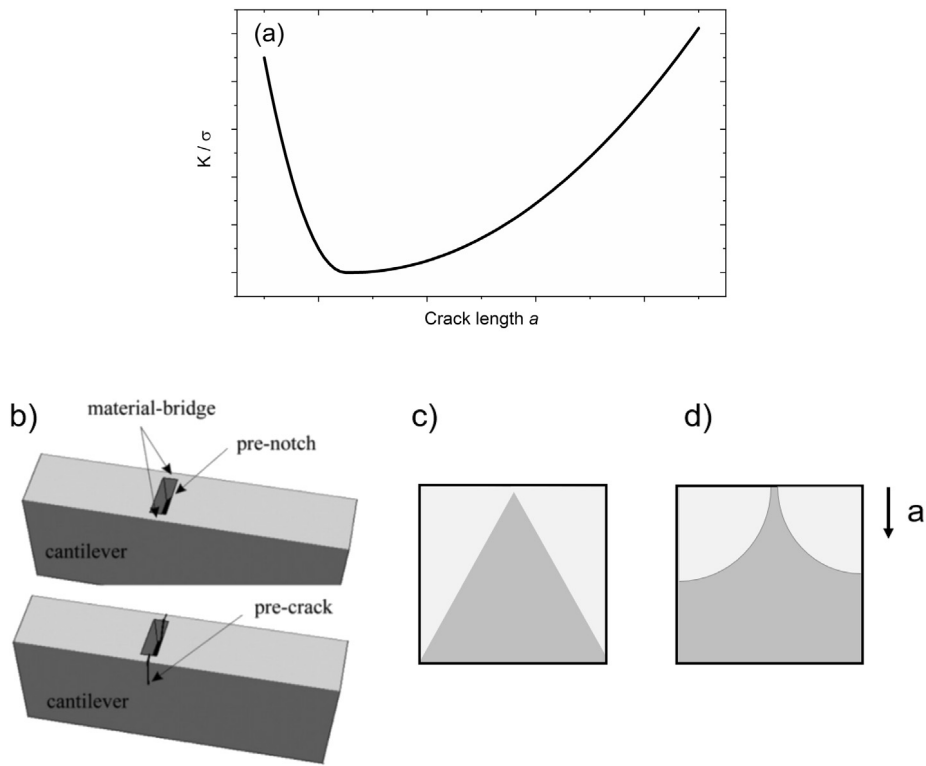

**Fig. 9.** Schematic illustration of the change of the normalized stress intensity factor as a function of crack length for chevron-like notches (a). The typical shapes of notch geometries to generate pre-cracks in ideal brittle material are depicted in (b), (c) and (d).

The criteria for different kinds of pore nucleation are not well understood. For modelling of fracture phenomena, usually the stress and/or plastic strain are used to determine the initiation of a crack. By varying sample size and the geometry, as well as the loading type (bending, tension or shear) a large variation of constraint and multiaxiality can be realized, and the conditions for the initiation of pores can be quantitatively evaluated.

Another important issue of micro ductile fracture is the growth of pores. The continuum mechanics description of the growth of voids in a triaxial stress field is well established [76]. This is not straightforward when extending this concept to nanopores and complex pore – microstructure interaction. Here, micro- or submicron samples can help to improve the understanding of the important physical phenomena during growth, for example for the analyses of the nanopore-dislocation interaction. Interesting questions which could be solved are: what is the critical stress and size for dislocation generation at pores, and how these generated pores affect nearby dislocations.

### 4.3. Semi-brittle fracture

Between the two extremes of ideal brittle and micro-ductile fracture, there is a very wide class of fracture phenomena, summarized as semi-brittle. Even in the case of micro-ductile fracture, the first stage of damage by pore generation is often a semi-brittle process. In contrast to micro ductile crack propagation, which is a consequence of the plastic coalescence of the pores, the semi-brittle fracture is a crack extension over small dimensions by a decohesion similar to ideally brittle crack propagation. In body centered cubic (bcc) metals or intermetallics below the ductile to brittle transition temperature, the decohesion along specific crystallographic planes or interfaces is associated with a certain level of dislocation or twining activity. The fracture toughness is therefore significantly higher compared to the ideally brittle case. Despite the technical importance of this fracture mode, the details of the underlying processes and the controlling parameters are not well understood.

Semi-brittle crack propagation is frequently observed in ceramics and many different composites, where additional irreversible processes such as phase transformation or wide-spread material damage takes place [66,68,70]. A further important feature of semi-brittle crack propagation is the formation of irregular, complex 3D shaped crack fronts by:

- Crack bifurcation
- Crack tunneling along specific crystallographic planes, resulting in the occurrence of non-connected facets
- Formation of a crack bridging zone
- Development of a region of micro-cracks
- Generation and extension of a single dominant micro crack in front of the blunted crack tip which results in failure of the whole sample
- And many other phenomena.

Hence, the variation of the fracture process zone extent during crack propagation can be exceptionally large, varying from few nm to m. Extreme examples are concrete or geological materials, where the size of the fracture process zone is equal to the extent of the non-linear deformed regime in front of the crack.

Consequently, sample size independent fracture toughness values can be expected only in materials with submicron or nanoscale microstructure where the fracture process zone is smaller than $r_K$ and $r_{HRR}$ of the micro sample. Furthermore, the microstructural length scale associated with the deformation should also be smaller than $r_K$ or $r_{HRR}$ (for examples see [74,77]).

Despite these limits in the determination of size-independent fracture resistance, miniaturized fracture mechanics experiments can be very helpful for improving the understanding of the intrinsic semi-brittle crack propagation processes. For example, the micron-sized samples permit significant reduction of the effect of crack bridging or microcrack shielding during semi-brittle crack propagation. Thus, the different contributions to the crack growth resistance can be discriminated. For example, the contribution to the global fracture resistance from decohesion or bridging can be elucidated with fracture mechanics microsamples from well-defined microstructural features, where the crack is aligned or inclined to cleavage planes or interfaces.

Furthermore, the sample size dependence of the plastic deformation resulting in a size dependent arrangement of dislocations in front of the crack tip offers completely new possibilities to study the fundamental dislocation-assisted decohesion processes. This will be discussed in more detail in Section 5.

The ability to test fracture mechanics specimens down to the sub-μm regime offers a vast number of additional opportunities to improve the current understanding of semi-brittle crack propagation. For example, in bcc metals, we have the ability to distinguish between propagation of the main crack or the formation of a micro-crack in front of the somewhat blunted main crack causing failure [69,78]. The latter is typical for ferritic steels near and below the ductile to brittle transition temperature, see for example [70]. The ability to vary sample size offers the possibility to avoid the formation of such a single micro-crack, or to significantly vary the stresses and strains where it originates. This can again serve as feedback for more damage tolerant designs.

Despite the inability for micro-scale samples to produce material characteristic fracture resistance measurements, these experiments will enable us to better elucidate individual processes operative during fracture. This will permit the development an understanding and methodology to predict the size dependence of the fracture resistance when the size-independent fracture mechanics description is not applicable.

### 4.4. Fatigue crack propagation

Under static conditions, the fracture toughness ($K_C$, $CTOD_C$, $J_i$, or $J_c$) determines the critical loading condition where cracks start to propagate and failure takes place for a crack-containing component, independent of whether this value is size dependent. In the sample size dependent case, both the size dependent behavior and the loading type (constraint effect) [10,28,29] have to be taken into account.

In a fatigue situation, similar to macro-scale components, the lifetime under cyclic loading of micro-components is determined by the generation of nano- or submicron cracks, the growth of already existing defects or newly generated cracks, and the final crack propagation that is determined by the fracture toughness.

For the applicability of fracture mechanics to the description of fatigue crack propagation in micro-devices or micro-samples, similar considerations as in the case of static loading have to be applied. Additional important length scales in fatigue are the extent of the cyclic plastic zone ($\Delta r_{pl}$), the cyclic crack tip opening ($\Delta CTOD$), and the cyclic fracture process zone ($\Delta r_{fr}$). Relevant quantities of fatigue crack propagation are given as follows [79]

$$\Delta r_{pl} = c_1 \frac{(K_{max} - K_{min})^2}{4\sigma_y^2} \tag{4.1}$$

$$\Delta CTOD = c_2 \frac{(K_{max} - K_{min})^2}{2E\sigma_y} \tag{4.2}$$

With $c_1$ and $c_2$ being constants depending on the material's hardening behavior and the stress state.

Depending on the applied loading conditions, these parameters can vary in a similar way as the corresponding quantities in the case of monotonic loading. However, for all common materials ΔCTOD and $\Delta r_{fr}$ are somewhat smaller than 1 nm at the onset of cyclic crack propagation, usually denoted as threshold ($\Delta K_{th}$ or $\Delta J_{th}$).

To assess whether such measured crack growth data are size dependent or not, one has to refer to the same considerations regarding length scales as in the case of monotonic loading. Regarding fracture processes, the materials can be again divided into the three previously mentioned classes:

- Ideally brittle materials, which only exhibit fatigue crack propagation in micro-samples due to some environment effects
- Semi-brittle materials and
- Ductile materials

The respective growth mechanisms for semi-brittle and ductile fatigue crack propagation are schematically depicted in Fig. 10.

In ductile materials, the fatigue crack propagation mechanism is significantly different from the monotonic case. The fatigue cracks grow by a blunting (where sometimes nano-cracking might be involved) and re-sharpening process [80–83]. The crack propagation rate is approximately proportional to $\Delta$CTOD, and the size of the fracture process zone is of the same order of magnitude [83,84]. However, this behavior is dependent upon the specific fracture process that occurs. In many polymers or biomaterials the ratio between $\Delta$CTOD, the fracture process zone, and the crack growth rate might be significantly different, as the molecular damage mechanism operative is much different than deformation in most crystalline structural materials.

Another important parameter in fatigue crack propagation is the so-called crack closure effect. During fatigue loading the crack is not open during the complete load cycle [85,86]. For a single parameter description of fatigue crack propagation an effective driving force ($\Delta K_{eff}$ or $\Delta J_{eff}$) is often used. In micro-samples, as in the case of short crack problems in macro-scale samples [49,68,87], the crack closure effect will be less important or negligible. Nevertheless, to compare fatigue crack growth data with macro-scale experiments it is essential to take into account the difference in the crack closure effect.

Although $\Delta$CTOD and $\Delta r_{fr}$ are in many cases small compared to the sample dimensions ($a$, $B$, $W-a$), the corresponding d$a$/d$N$ vs. $\Delta J_{eff}$ curve will be sample size dependent. Only when the microstructural dimensions are small compared to $\Delta r_{HRR}$ can a size independent cyclic crack growth d$a$/d$N$ vs. $\Delta J_{eff}$ be expected, because then the cyclic plastic deformation near the crack tip will not depend on the geometry or dimensions of the sample. Comparison of d$a$/d$N$ vs. $\Delta$CTOD data of long and short cracks in macro-samples are in good agreement [83], i.e. the typical short crack effects disappear. Similar behavior is also expected in micro-samples. In ductile metals, where $\Delta r_{fr}$ is about $\Delta$CTOD, the relation between d$a$/d$N$ and $\Delta$CTOD is expected to be sample size independent in a relatively large loading regime between the threshold and $\Delta$CTOD of ~1/20 of $B$, $a$ and $W-a$. This limit of size independence is equal to the standard dimensional criterion for a critical CTOD determination [67].

## 5. Dislocation arrangements near the crack tip

In order to visualize the effect of crack length on the dislocation arrangement, the linear elastic stress field in a macroscopic tension sample with a through-thickness crack with lengths from nm to mm is schematically depicted in Fig. 11. The assumed stress intensity factor is small, 0.5 MPa m$^{1/2}$, a typical value where dislocations can be generated at crack tips in metals [34]. Taking into account the typical friction stress of a dislocation of a few MPa, it is evident that for long cracks, dislocation movement is restricted by the K-dominated elastic stress field. For crack lengths in the micrometer regime, the movement of the first generated dislocations in coarse grained metals will most likely not be restricted by the K-dominated field, but by microstructural barriers such as grain or phase boundaries. This is because the applied stress necessary for a stress intensity factor sufficient to generate a dislocation at the crack tip is relatively large compared to the friction stress, while the K-dominated region is very small.

| | Size of the fracture process zone in fatigue |
|---|---|
| **Ductile Material** | |
| [schematic load-time plot and crack tip geometry diagram] | ~ $\Delta$CTOD = 3 Å – 100 µm |
| **Semi-brittle material** | |
| [SEM micrograph, scale bar 50 µm] | 3 Å – mm |

Fig. 10. Illustration of the fatigue crack propagation mechanism in ductile and semi brittle material and the corresponding extent of the fracture process zones. For the ductile material, the cross-sectional geometry of the crack tip and its change within a single loading cycle (steps 1 to 6) is depicted. Actual crack extension is shown in red color. For the semi-brittle material, a typical fracture process involving cleaving and crack bridging is shown.

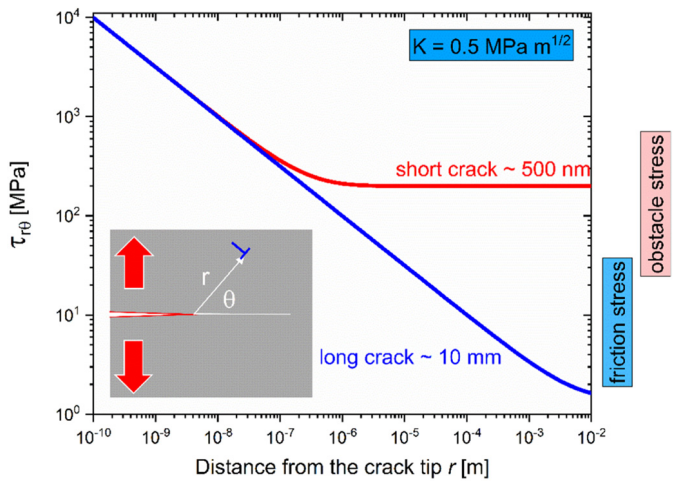

**Fig. 11.** Illustration of the shear stresses which act on the first dislocation generated from the crack tip for a tension loaded sample with different crack lengths. The typical friction and obstacle stresses are also indicated.

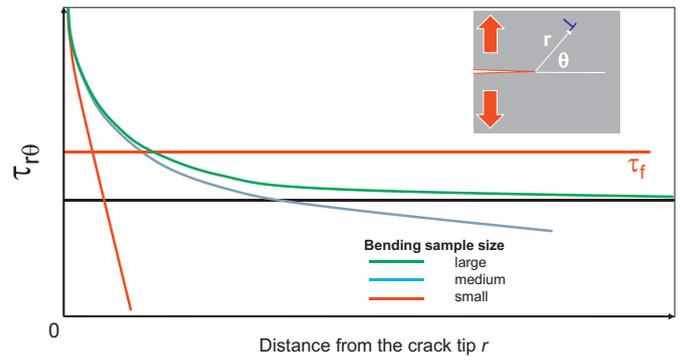

**Fig. 12.** Schematic illustration of the shear stress acting on the first dislocation generated from the crack tip in bending samples with different ligament lengths loaded by the same stress intensity factor. For large specimens, the shear stress controlling the dislocation motion is solely governed by crack stress field, whereas for small specimens, the proximity of the neutral axis leads to a more rapid decline of the applied shear stress. The typical friction stress for single crystal micro-samples are indicated in order to illustrate the strong sample size effect in the dislocation arrangement in bending samples. The dimension of the large bending specimen exceeds the depicted regime.

For very small cracks (few nm) the necessary stress to generate a dislocation at the crack tip becomes very high, significantly larger than the typical yield stress of a coarse grained metal or alloy. Hence, other dislocation sources at large distances from the crack will be active at significantly lower stresses than that required to generate a dislocation at the crack tip. Since the K-dominated regime of a few nm long crack is small (few nm), the plastic deformation will not be affected by a tiny crack in a typical low and medium strength material.

Changing both the crack length and the ligament width ($W-a$) of a cracked tensile sample by the same order of magnitude, i.e. down to the submicron regime, does not change the elastic stress field significantly. The only difference is that the "far-field" stress will only slightly increase. However, when the sample size is reduced to the nm regime, the original polycrystalline experiment is reduced to a single crystal experiment and the probability of finding dislocation sources will be significantly reduced.

Furthermore, since typical obstacles for dislocation motion within such micro-samples often disappear, a generated dislocation at or near the crack tip can escape from the sample. Hence, the generated arrangement of dislocations will be significantly different in such micro-tension fracture samples compared to macro-scale samples at the same CTOD (i.e. the same amount of plastic deformation of the crack tip). Dependent on the crack length, in the miniaturized fracture mechanics tensile samples the dislocations will be more loosely packed compared to macro-scale samples.

The situation is opposite in micro-bending samples. Micro-cracks in macroscopic bending samples behave very similarly to those in macro-scale tension samples. In single crystalline fracture mechanics micro-bending samples, however, the dislocation arrangement will be completely different compared to the micro-tension condition due to the dislocation pile-up at the neutral axis [18]. This difference is clearly evident from the analysis of the linear elastic stress fields experienced by the first generated dislocation. In large bending samples the first generated dislocations are governed by the K-dominated field, whereas in micro-samples the dislocation arrangement will be controlled by the dislocation pile-up at the neutral axis of the bending sample [88,89]. The extent of dislocation pile-up will not only be dependent on the sample size (crack length, ligament length) but also on the orientation of the dislocation glide plane(s) within the sample. Stresses acting on dislocations within micro-bending samples of different dimensions loaded with the same stress intensity factor are schematically shown in Fig. 12.

Hence, in micro-bending, the dislocations in front of the crack will be even more densely packed compared to macro-scale samples, and the reduction of the bending sample size into the micron and submicron regimes dramatically changes the dislocation arrangement in front of the crack tip. One may expect that this effect will induce significant sample size dependence in the fracture resistance and increase the difficulty in predicting the fracture of micro components. For the understanding of semi-brittle fracture processes (crack propagation by cleavage of crystals along well defined planes, grain boundaries or phase boundaries) these strong but controllable changes of the dislocation arrangements at the crack tip will open completely new ways to answer basic questions in this important area of fracture mechanics of materials.

## 6. Relevance of plane strain versus plane stress boundary conditions

In fracture mechanics, plane deformation states are usually considered, i.e. plane stress or plane strain. The reason for this classification is that only for these two extreme cases do analytical solutions for the relation between applied loading and the energy release rates G, K, and J exist [10]. Even more importantly, the solutions for the near crack tip stress field are only valid for the two extreme cases. Because the thickness of micromechanical samples is in the μm regime, a plane stress behavior is frequently assumed, but this is correct only in selected cases.

Dependent upon the problem – conversion of K into G or J, consideration of the fracture process itself, or formation of the plastic zone – the phenomena should be better approximated by the plane stress or the plane strain case. Figs. 13 and 14 show schematically the important differences and the regimes where a plane stress or plane strain approach are useful, respectively.

In the linear elastic case, the stresses $\sigma_{11}$ and $\sigma_{22}$ (for notation see Fig. 1) are not affected by the stress state; only $\sigma_{33}$ and the resultant displacement field differ between plane stress and plane strain conditions (see for example [10,26]).

In case of an elasto-plastic behavior in the vicinity of the blunted crack the stresses differ significantly between the two stress states. In plane strain, the stresses $\sigma_{22}$, $\sigma_{33}$, $\sigma_{11}$ in and ahead of the fracture process zone can reach about 3, 2.5, 2 times the flow stress of the material, respectively. In plane stress, $\sigma_{22}$ is about equal to the flow stress $\sigma_y$.

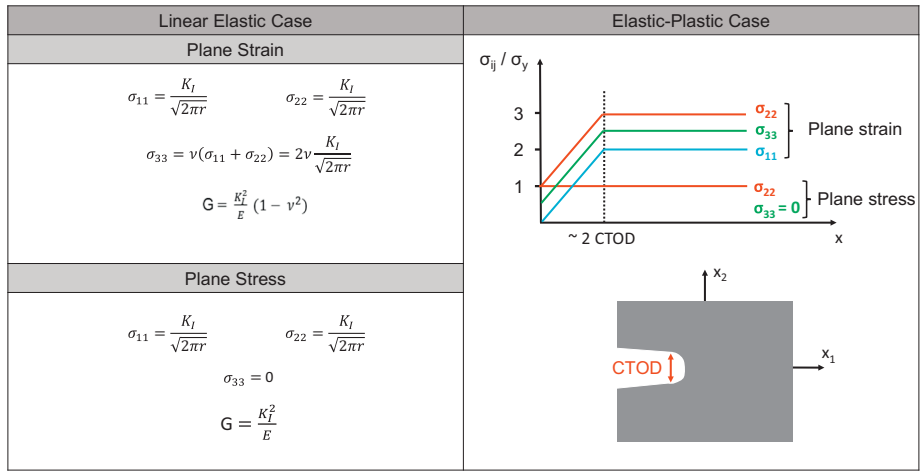

**Fig. 13.** Comparison of the principle stresses along the $x_1$ axes in front of a mode I loaded crack for the ideally linear elastic case and in the plastic zone for elastic perfectly-plastic material for the plane stress and plane strain case (the stresses are normalized by the yield stress).

For the conversion of K to G in small-scale yielding, the plane stress approach is a good approximation if the thickness of the sample is smaller than $r_K$. Hence, for thin plates with crack lengths much larger than the plate thickness, the plane stress conversion of G to K is appropriate. For the standard fracture mechanics toughness samples where crack length *a*, thickness *B* and ligament (*W-a*) are about equal, the plane strain approximation is the better choice.

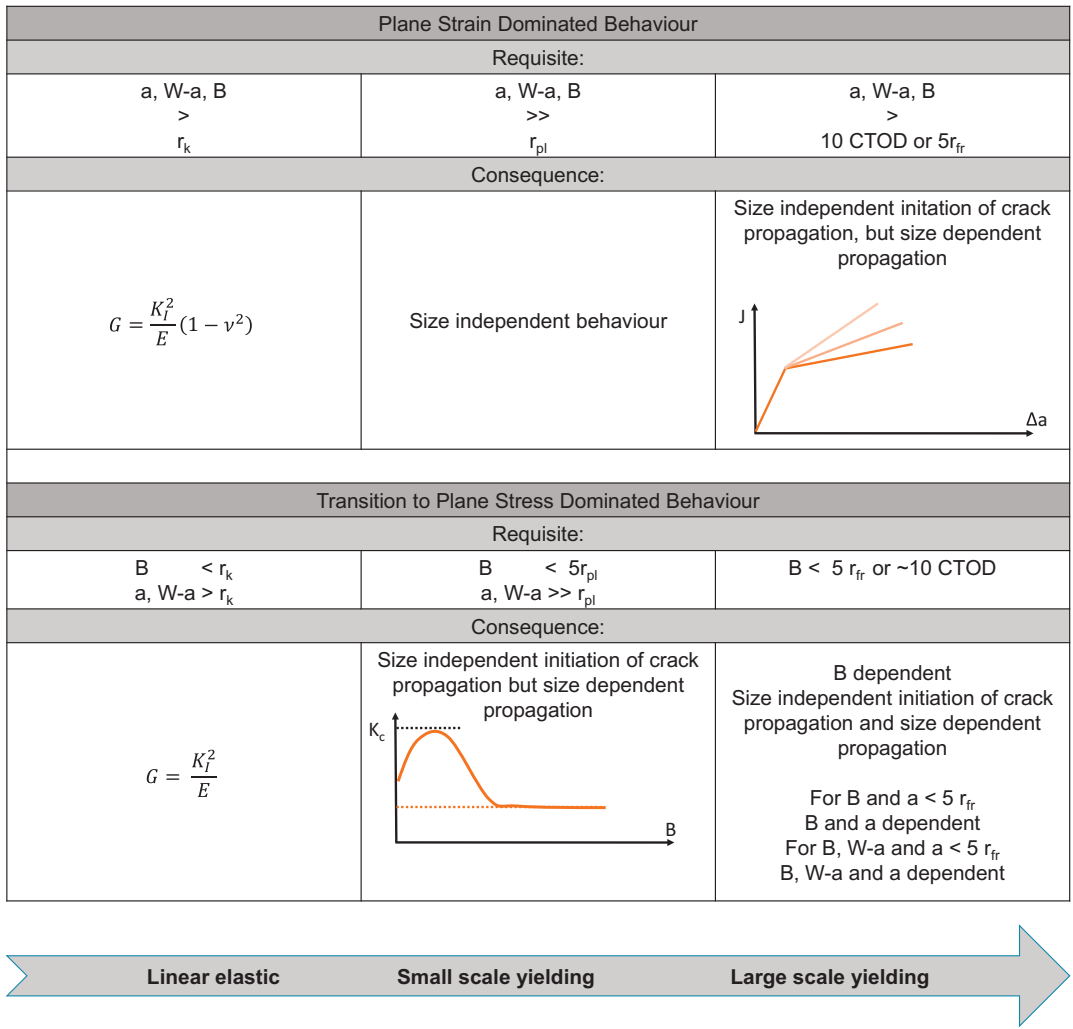

**Fig. 14.** Illustration of the different loading regimes of cracked specimens, and the geometry and size dependence on the fracture resistance for plane strain and plane stress dominated deformation behavior.

For the fracture resistance of a material, the stress state inside the plastic zone and the fracture process zone are more important than the stress state in the surrounding elastic region. The plane strain state dominates the shape and the deformation in the plastic zone if the sample thickness is large compared to the plastic zone size, which is well fulfilled if $B > 10 r_{pl}$. In this case, the determined critical K value can be considered as a material characteristic parameter, the plane strain fracture toughness, $K_{Ic}$.

If this is not the case, the plastic deformation in the plastic zone becomes more and more plane stress dominated. This change induces an R-curve behavior of the crack propagation resistance. Such behavior can be an intrinsic material effect, which is the case in many semi-brittle metallic, intermetallic, or ceramic materials [66]. If the fracture process zone is smaller than $r_K$, such an R-curve is sample size independent. Another reason for such R-curve behavior is the transition from the plane strain dominated deformation in the plastic zone to a plane stress affected deformation mentioned above. In such case the apparent fracture toughness becomes thickness dependent. It increases with decreasing sample thickness, as long as the fracture process zone is small compared to the sample thickness. However, in the center of the specimen the first crack propagation starts at a stress intensity or a J-value, which is equal to the plane strain fracture toughness. The increase of the fracture resistance is in this case a consequence of the plane stress dominated fracture process in the near surface regime of the sample. The R-curve is only thickness independent if small scale yielding is fulfilled [90], i.e. *a* and *W-a* is significantly larger than $r_{pl}$ and $r_{pl} < r_K$, except the thickness criterion.

If the fracture process zone under large scale yielding condition is mainly controlled by plane strain deformation, the J-integral or CTOD can be used to describe the fracture resistance. In standard macro-scale experiments the initiation toughness ($J_i$ or $CTOD_i$) is usually geometry independent, as long as certain specimen size requirements are fulfilled. However, the crack growth part of the J versus Δ*a* or CTOD versus Δ*a* curve are sample size dependent even in macro-scale samples [91–93]. In micro-samples $J_i$ and $CTOD_i$ are only size and loading type independent if the plastic deformation is not specimen size or loading type dependent. This is typically the case only when the characteristic length scale of the microstructure is very small compared to the sample size.

If the extent of the fracture process zone approaches the sample thickness, or *B* is smaller than 10 $CTOD_i$, a sample size dependent initiation of crack propagation as well as a size dependent propagation is expected. In such thin, ductile samples, the fracture resistance usually decreases with decreasing sample thickness, which is typical in ductile thin free standing films [94]. Crack propagation in thin films is very important in a vast number of engineering applications. This problem has to be subdivided into different subjects regarding the type of films (free-standing or supported films, compliant or stiff substrates, rigid or plastically deformable substrates), and the particular fracture process. In addition, one has to distinguish between crack propagation behavior of the film material itself, and the possible decohesion process from the substrate in an ideally brittle, semi-brittle or ductile failure processes. This area of fracture mechanics is an exceptionally wide research field. For more information, the reader is referred to the following review [95]. Despite some similarities with discussed problems in fracture mechanics for samples or components where all dimensions approach the micron or submicron regime, this will not be discussed in this paper in favor of keeping a clear focus.

## 7. Applicability of macroscopic fracture mechanics test standards

As long as small scale yielding conditions are fulfilled, i.e. $r_{pl} < r_k$, the standards for linear elastic fracture mechanics for fracture toughness, fatigue crack propagation or environmental assisted crack propagation are applicable. However, this conditions can be rarely achieved in micro-sized fracture mechanics samples except for ideal brittle materials or high strength materials with low fracture toughness. This is a common case in nanostructured materials, especially for materials and structures in devices with functional application. Once the characteristic microstructural dimensions are not small compared to the sample dimensions, the fracture resistance measured will be sample size dependent even if small scale yielding is fulfilled. For such case the arrangement of the microstructure in relation to the crack and sample geometry has to be taken into account, the resulting values being quite structure sensitive.

Since in most cases, even in relatively brittle materials or near the threshold for fatigue crack propagation, small-scale yielding conditions cannot be fulfilled in micrometer sized specimens, elasto-plastic fracture mechanics concepts (J, ΔJ, CTOD or ΔCTOD) have to be used to characterize the fracture resistance. Even if CTOD meets the condition for predominantly plane strain deformation at the crack tip (in the mid-section of the sample), the determined values might be microstructure-, sample size-, or loading type- dependent, as discussed above.

Due to the difficulties and uncertainties in the determination of a critical crack initiation $J_i$ (the J-integral value where blunting ends and stable crack propagation starts), $J_{0.2}$ is usually used as fracture toughness value for macroscopic specimens. This $J_{0.2}$ is the J-value for 0.2 mm crack extension after the blunting. Since sample sizes of micro-samples are typically smaller than 0.2 mm, such standard $J_{0.2}$-evaluations are not applicable [96].

The stable crack propagation in the J versus Δ*a* plot is denoted as the tearing regime and the slope shows the tearing modulus. The tearing modulus is sometimes assumed to be a material characteristic property, however, that is not correct. The slope contains always a certain sample size dependence [91–93]. A very simplified view of this dependence can be obtained from the schematic picture in Fig. 15. Following the arguments of Irwin and Orowan [2,3] used to explain the fracture toughness under small scale yielding in ductile materials, the necessary energy to create a fracture surface is given by 2($\gamma_0 + \gamma_{pl}$), where $\gamma_0$ is the surface energy and $\gamma_{pl}$ the plastic work to generate the fracture surface. $\gamma_{pl}$ can be divided into the work to form the micro-ductile dimples $\gamma_{pl\ fr}$ and the work to move the plastic zone through the sample $\gamma_{pl\ plz}$ (Fig. 15). As long as $r_{pl} < r_k$, $\gamma_{pl\ plz}$ does not depend on the sample dimension. However, in the case of EPFM ($r_{pl} > r_k$), $\gamma_{pl\ plz}$ will be specimen size dependent.

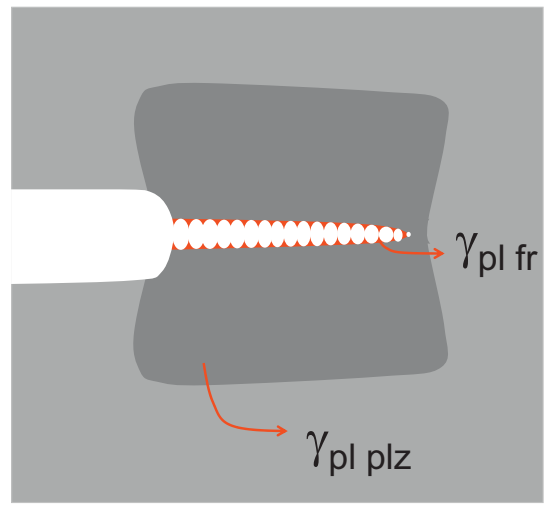

**Fig. 15.** Schematic illustration of the plastic work spent during micro ductile crack propagation in the case of small scale yielding. The plastic work can be separated into the work necessary for the forming of the dimple-like fracture surface and that to move the plastic zone through the specimen.

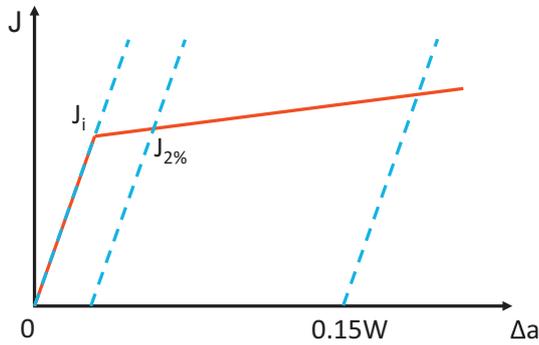

**Fig. 16.** Illustration of the J versus Δa curve adjusted to micron or submicron fracture mechanics samples. Determination of the tearing modulus (dashed lines) of the macroscopic standard J approach is adapted to micro samples. Thus, $J_{2\%}$ instead of $J_{0.2mm}$ is evaluated.

For the evaluation of the specimen size independent $J_0$ or $J_{0.2}$ in macro-scale samples, a linear or a power law fit in the regime between 0.15 and 1.5 mm of physical crack extension (not including the extension by blunting) is used to determine the intersection with the blunting line or the 0.2 mm offset of the blunting line [67]. A possible transfer of this concept to micro-samples could be the determination of a $J_i$ or $J_{2\%}$ by intersecting the blunting line or an offset of the blunting line by 2% of the crack length or 2% of the sample width with a linear fit of the tearing regime in the crack extension interval Δa between 1% and 10% of the crack length or the sample width, W, as shown in Fig. 16. Procedures following this concept or targeting $J_i$ are being established [97,98].

If the sample size for the $J_i$ or $J_{2\%}$ values meets the standard dimensional criterion, the values determined are material characteristic only if the characteristic microstructural dimensions are small compared to $r_{HRR}$. In the other cases, the values are characteristic for a certain microstructure, sample size and loading type. If the specimen size criterion of $J_i$ or $J_{2\%}$ is not fulfilled, the values are a-priori size dependent.

Furthermore, it should be noted that the blunting line is a consequence of continuum plasticity (i.e. a sample size and gradient independent yield stress). However, at very small crack tip opening displacements, continuum plasticity overestimates the plastic deformation. Due to the discrete nature of plasticity and the sample size dependence, in the nm regime the "blunting line" should not be a perfect line [42]. Moreover, since many of the micro-mechanics fracture experiments are performed in situ in an SEM, the presence of vacuum has to be taken into account. For static loading this effect should be of minor importance compared to an inert environment, however, in some materials even humid air may affect the crack propagation behavior. In case of cyclic loading, the effect of vacuum can be more pronounced [87]. However, if the loading frequency is not too high and standard medium vacuum conditions are used, the effect should be small. As a rule-of-thumb, at an oxygen pressure of $10^{-6}$ mbar in 1 s an oxide monolayer is formed. Hence, in typical low frequency fatigue experiments in an SEM, in each loading cycle an oxide layer can be formed.

## 8. Final remarks

As already mentioned, the new micromechanics possibilities of studying the behavior of well-defined cracks in miniaturized samples opens enormous possibilities for the development of damage tolerant designs of micro-devices, as well as for the understanding of crack propagation in general. Some emerging possibilities to address long-standing issues will be mentioned in the following to inspire possible future research directions.

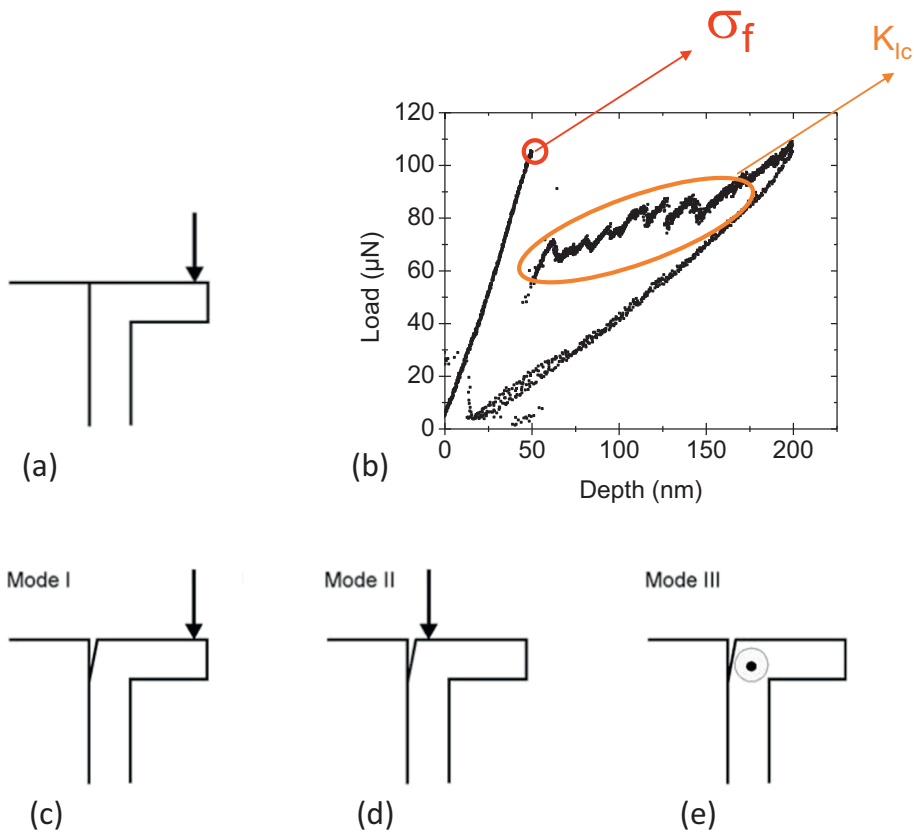

**Fig. 17.** Illustration of a simple procedure to measure a) the ideal strength of weak interfaces, and the fracture toughness in c) mode I, d) mode II and e) mode III using a single specimen type. b) An example presents the load displacement curve to determine the theoretical strength and the mode I fracture toughness in such weak interface [60].

In the last few years, techniques have been developed for in situ study the reaction of a material to aggressive environment, even in a TEM. A combination of such capabilities, often referred to as *in operando*, with micro/nano fracture mechanics experiments will open new ways to solve important questions in the challenging area of stress corrosion cracking and fatigue corrosion cracking.

Furthermore, using the sample size and loading type dependence of deformation and dislocation arrangement, additional insights in the environmentally assisted crack propagation can be obtained (for example relating to the phenomena of hydrogen embrittlement [99]). In the case of environmentally assisted crack propagation, in addition to the mentioned characteristic lengths, the extent of the environmentally affected region has to be taken into consideration.

In most of the studies performed (see [21]) crack propagation under predominantly mode I loading was assumed/performed. The important length scales in micro-fracture mechanic experiments are not limited to mode I loading. Mixed mode failure is a quite common phenomenon in micro devices. Hence, the analysis of mode II and III and the effect of mode mixing will be an important future topic.

As in the case of mode I loading, such mode II and III will be very important for the better understanding of crack propagation in general. In macro-scale experiments the crack flank interactions – interlocking and friction – induce enormous uncertainties in the interpretation of the observed behavior. This uncertainty can be avoided or be significantly reduced in micro samples. Fig. 17 illustrates schematically the simple possibilities to generate strength and fracture toughness data of interfaces in micro samples for different loading modes [60]. However, it should be mentioned that for the quantitative evaluation of the interface properties a careful simulation of the experiment is required.

## 9. Conclusion

The crack growth resistance of micrometer and sub-micrometer sized samples is considered from the fracture mechanics point of view. Special attention is devoted to the different relevant length scales which should be taken into account in such experiments.

The most important parameters are the dimensions of the fracture process zone in relation to the fracture mechanics parameters, the extent of the K dominated regime, the magnitude of the HRR-field, the crack tip opening (or shear) displacement, and the structural length scale of the material (grain size, dislocation spacing etc.). For each combination of material properties and microstructural sizes, these parameters have to be considered and specimen dimensions and validity of experimental results have to be critically considered. Owing to the complexity of the fracture mechanical problems, a short summary – or even better a process flow chart – is very difficult to be presented. Intrinsic material characteristics, and therefore sample size independent crack propagation resistance ($K_{IC}$, $J_{IC}$, $CTOD_C$, $da/dN$ versus $\Delta K$, or $da/dt$ versus $K$), can only be determined on such small specimens in ideally brittle situations or for materials with low fracture toughness and very high strength.

In many cases, the fracture resistance in micro-sized components or specimens will be sample size dependent because the required dimensional relations are not fulfilled. The determination of this specimen size dependence is an essential future task to enable damage tolerant design and life time prediction of nano- or micro-devices. In addition, these miniaturized fracture mechanics experiments will open completely new ways to better understand the fracture processes taking place in macroscopic materials.

## Author contribution

R.P. wrote the main body of the manuscript, basing on joint ideas. S. W. and D. K. complemented and completed the manuscript.

## Acknowledgement

This project has received funding from the European Research Council (ERC) under the European Union's Horizon 2020 research and innvocation programme (ERC Grant Agreement No. 340185 USMS, ERC Grant Agreement No. 757333 SpdTuM, and ERC Grant Agreement No. 771146 TOUGHIT.)

The authors thank Kurt Matoy for performing the work and providing the data for the schematic overview on mode II and mode III loading, and Glenn Balbus for careful manuscript editing.